

\input harvmac

\def\LG{Lan\-dau-Ginz\-burg\ }
\def\BW{Boltz\-mann weights\ }

\def\half{{1 \over 2}}

\def\annp#1{{\it Ann. \ Phys.}\ {\bf #1\/}}
\def\cmp#1{{\it Commun.\ Math. \ Phys.} \ {\bf #1\/}}
\def\nup#1{{\it Nucl.\ Phys.} \ {\bf B#1\/}}
\def\plt#1{{\it Phys.\ Lett.}\ {\bf #1\/}}

\def\jsp#1{{\it J. \ Stat. \ Phys.}\ {\bf #1\/}}

\def\mpl#1{{\it Mod.\ Phys.\ Lett.} \   {\bf A#1}\ }

\def\scf{{\cal F}}
\def\scp{{\cal P}}
\def\scq{{\cal Q}}
\def\Gminus{G_{-{1 \over 2}}^-}
\def\Gplus{G_{-{1 \over 2}}^+}

\def\coeff#1#2{\relax{\textstyle {#1 \over #2}}\displaystyle} 
\def\inbar{\vrule height1.5ex width.4pt depth0pt}
\def\IC{\relax\,\hbox{$\inbar\kern-.3em{\rm C}$}}
\def\IR{\relax{\rm I\kern-.18em R}}
\font\sanse=cmss12
\def\ZZ{\relax{\hbox{\sanse Z\kern-.42em Z}}}

\font\ninerm=cmr9

%
\def\Titletwo#1#2#3#4{\nopagenumbers\abstractfont\hsize=\hstitle
\rightline{#1}\rightline{#2}
\vskip .7in\centerline{\titlefont #3}
\vskip .1in \centerline{\abstractfont {\titlefont #4}}
\abstractfont\vskip .5in\pageno=0}

\def\Date#1{\leftline{#1}\tenpoint\supereject\global\hsize=\hsbody%
\footline={\hss\tenrm\folio\hss}}
%

\Titletwo{}{}
{Off-Critical Lattice Analogues of $N=2$ Supersymmetric}{Quantum Integrable
Models$^*$} {} {}
\centerline{D. Nemeschansky \ and \ N.P. Warner}
\bigskip \centerline{Physics Department}
\centerline{University of Southern California}
\centerline{University Park}
\centerline{Los Angeles, CA 90089-0484.}
\vskip 1.0cm
We obtain off-critical (elliptic) Boltzmann weights for lattice models whose
continuum limits correspond to massive, $N=2$ supersymmetric, quantum
integrable field theories.  We also compute the free energies of these
models
and show that they are analytic in the region of parameter space where we
believe that the supersymmetry is unbroken.  While the supersymmetry is not
directly realized on the
lattice, there is still a very close connection between the models described
here and topological lattice models.  A simple example is discussed in
detail
and some corner transfer matrix computations are also presented.

\vskip .1in
\vfill
\leftline{USC-93/018}
\leftline{hep-th/9307141}
\Date{July 1993}
%


\newsec{Introduction}

\nref\EMa{E.~Martinec, \plt{217B} (1989) 431.}
\nref\VW{C.~Vafa and N.P.~Warner, \plt{218B} (1989) 51.}
\nref\EMb{E.~Martinec, {\it ``Criticality, catastrophes and
compactifications,''}  V.G. Knizhnik memorial volume, L.~Brink {\it  et al.}
(editors): {\it Physics and mathematics of strings.} }
\nref\LVW{W.~Lerche, C.~Vafa and N.P.~Warner, \nup{324} (1989) 427.}
\nref\NPW{N.P. Warner, {\it ``Lectures on N=2 superconformal theories and
singularity theory'',in ``Superstrings '89,''} proceedings of the Trieste
Spring School, 3--14 April 1989.  Editors: M.\ Green, R.\ Iengo,  S.\
Randjbar-Daemi, E.\ Sezgin and A.\ Strominger.  World Scientific (1990);
{\it
N=2 Supersymmetric Integrable Models and Topological Field Theories ,}
Lectures given at the Summer School on High Energy Physics and Cosmology,
Trieste, Italy, June 15th -- July 3rd, 1992.  To appear in the proceedings.}

In studying $N=2$ superconformal models, and their perturbations, the \LG
approach has proved to be extremely powerful.  The basic idea of this
approach
is to isolate a special subset of the fields, the ``chiral primaries,'' and
to
try to describe the entire model in terms of an effective field theory of
this
subset of fields \refs{\EMa {--} \NPW}.   The term ``\LG formulation'' has
thus
taken on a slightly broader meaning than that of finding an effective field
theory of order parameters of a statistical system.  However, the question
naturally arises as to what extent the sense of having a ``\LG formulation''
has shifted,  that is, one would like to know if one can, in fact, find
lattice
models whose natural order parameters are the \LG fields of  $N=2$
supersymmetric models.

Lattice analogues\foot{We use the terminology ``lattice analogues'' to
connote the fact that most of these lattice models do not explicitly
realize the
supersymmetry on the lattice.  The supersymmetry thus may only be present in
the continuum limit.}  of many $N=2$ superconformal models have been
constructed.  The $N=2$
superconformal minimal series can be obtained by going to a special point in
the parameter space of the $k$-fused six vertex model.  This fact has been
employed in the recent work relating polymers and percolation to $N=2$
superconformal models \ref\HS{H.~Saleur, \nup{382} (1992) 486; H.~Saleur,
\nup{382} (1992) 532;  P.~Fendley and H.~Saleur, \nup{388} (1992) 609. }.
More
recently, it has been shown how to construct the {\it critical\ } Boltzmann
weights for exactly solvable lattice models whose continuum limits yield the
$N=2$ superconformal coset models based on hermitian symmetric spaces
\ref\MNW{Z.~Maassarani, D.~Nemeschansky and N.P.~Warner, \nup{393} (1993)
523.}.  It was also argued in \MNW\ that the order parameters of these
lattice models were indeed the \LG fields described in \LVW.  The method
employed in \MNW\ will be described briefly in the next section, but it
essentially employs a simple modification of the quantum group truncation
procedure in  certain classes of vertex models, or equivalently, a simple
modification of the height restrictions in the IRF formulation of the
lattice
models.  The arguments that such modifications lead to the lattice analogues
of
$N=2$ superconformal models were largely based upon the Coulomb gas
description
of the field theory and its correspondence with the lattice description.
More
recently these arguments have been further substantiated by Bethe Ansatz
calculations
\ref\ZMa{Z.~Maassarani, ``Conformal Weights via Bethe Ansatz for $N=2$
Superconformal Theories,'' USC preprint USC-93/019.}.

\nref\FMVW{P.\ Fendley, S.\ Mathur,
C.\ Vafa and N.P. \ Warner, \plt{243B} (1990) 257.}
\nref\FLMW{P.~Fendley, W.~Lerche, S.D.~Mathur and N.P.~Warner, \nup{348}
(1991) 66.}
\nref\PMMAW{P.~Mathieu and M.A.~Walton \plt{254B} (1991) 106.}
\nref\Grisaru{M.T.~Grisaru, S.~Penati and D.~Zanon, \plt{253B} (1991) 357;
G.W.~Delius, M.T.~Grisaru, S.~Penati and D.~Zanon, \plt{256B} (1991) 164;
\nup{359} (1991) 125.}
\nref\Hollow{J.~Evans and T.J.~Hollowood \nup{352} (1991) 723;
\nup{382} (1992) 662; \plt{293B} (1992) 100.}
\nref\NemWar{D.~Nemeschansky and N.P.~Warner, \nup{380} (1992) 241.}

One of the purposes of this paper is to give elliptic Boltzmann weights
that satisfy the Yang-Baxter equations, and whose critical limits yield the
critical Boltzmann weights given in \MNW.  These weights are given in
section 2, while section 3 describes a simple example.  We believe that
these models are lattice analogues of massive $N=2$ supersymmetric quantum
integrable models.  That the continuum limit is quantum integrable follows
from the exact solvability of the lattice model; it is the supersymmetry
that
is not so obvious.

There are several reasons for believing that at least the continuum limits
of
the off-critical lattice models are $N=2$ supersymmetric.  First,
at criticality in the Coulomb gas description,
there are natural lattice operators that correspond to
the perturbations that are known to lead to $N=2$ supersymmetric quantum
integrable models \refs{\FMVW {--} \NemWar}.  These lattice operators extend
the
quantum group structure to an affine quantum group \MNW\ and
this
fact leads us to believe that these operators will play the role of the
off-critical lattice perturbations.

\nref\Spieg{M.~ Spiegelglas \plt{274B} (1992) 21; M.~ Spiegelglas and
S.~ Yan\-kie\-lowicz \nup{393} (1993) 301.}
\nref\wittop{E.~Witten, \cmp{117} (1988) 353; \cmp{118} (1988) 411;
\nup340 (1990) 281.}
\nref\EY{T.~Eguchi and S.-K.~Yang, \mpl4 (1990) 1693.}

Another piece of evidence for the supersymmetry is the extremely close
relationship between our lattice models and rigid topological lattice
models.
In the continuum, the topologically twisted versions of the $N=2$
supersymmetric quantum integrable models are precisely the topological
$G_k/G_k$ models \refs{\Spieg,\NemWar}.  Under the twisting,
one of the $N=2$ supersymmetry
generators goes to the BRST operator that is
responsible for the truncation to the topological theory
\refs{\wittop,\EY}.  Our approach
is to reverse this course and consider lattice analogues of theories with
a topological $G/G$ subsector and then untwist them into new, non-trivial
lattice models.  We believe that our lattice construction is precisely the
discrete version of the continuum correspondence, and in particular, the
topological lattice model must have some implicit knowledge of the
topological
BRST operator, and that the process of ``untwisting'' will convert this into
a supersymmetry.  Similar observations have been made for geometrical
lattice
models \HS.

\nref\Bax{R.~Baxter, \annp{76} (1973) 1; \annp{76} (1973) 25;
\annp{76} (1973) 48.}
\nref\Baxter{R.J.~ Baxter, {\it Exactly Solved Models in
Statistical Mechanics,} Academic Press, London 1982.}
\nref\ABF{ G.E.~Andrews,  R.J.~Baxter and P.J.~Forrester,
\jsp{35} (1984) 193.}

In sections 4 and 5 we will amass further evidence for the supersymmetry of
our models.  In section 4 we compute the free energy of a class of models
that
includes our putative supersymmetric models.  We show that the
supersymmetric
models have exactly the same free energy as the corresponding {\it rigid}
topological lattice models, and that this free energy is analytic in the
regime where we believe that the continuum limit is supersymmetric.  As we
will
discuss, such analyticity of the free energy is certainly a necessary signal
of the supersymmetry \HS,
but it by no means proves that the model is supersymmetric.
It has long been known that the eight vertex model has analytic free
energy in regime $I\! I\! I$ for certain choices of one of the parameters
\refs{\Bax {--} \ABF}.  However only one of these choices leads to
a supersymmetric continuum limit.
We will discuss this further, along with some
generalizations, in section 4.

The most direct evidence for the supersymmetry is also given in section 4.
We
use hyperscaling arguments to compute the dimension of the perturbing
operator that takes the critical lattice model to the off-critical model.
Needless to say, the result is completely consistent with our identification
of the continuum limit.  In addition to this we describe how to smoothly
untwist the topological lattice model, and we examine how the dimension of
the perturbing operator changes under this process.  The result is not only
exactly parallel to the continuum $N=2$ supersymmetric theory,
but also enables us to read off the $U(1)$ quantum number of the perturbing
operator.  Again, the result consistent with our identification
of the continuum limit as an $N=2$ supersymmetric quantum integrable model.

Section 5 contains some results of a corner transfer matrix
computation.  As one would expect, we find that, at least for the simple
example given in section 3, the local height probabilities can be written
in terms of the characters of the $N=2$ superconformal models.  Our
computations also lead to the identification of $N=2$ \LG fields with the
order parameters.  In this section
we will also describe a slight refinement of the corner transfer matrix
computation of one-point functions.  This refinement generates to the $N=2$
superconformal characters replete with the gradation by $U(1)$ quantum
numbers.

Finally, there is the question of generalizations of our methods.  There are
certainly obvious ones: fusions of our models, other groups, other amounts
of supersymmetry, and even generalized supersymmetry \ref\ALCV{A.~LeClair
and
C.~Vafa,  {\it ``Quantum Affine Symmetry as Generalized Supersymmetry,''}
Cornell preprint CLNS-92-1150, hep-th@xxx.lanl.gov 9210009.}.  We will make
further remarks about these issues in the course of the paper or in the
final section.

\newsec{The Boltzmann Weights}

The key idea in obtaining the critical Boltzmann weights for $N=2$
superconformal coset models is to employ the connection with topological
$G/G$
models.  Specifically, consider the $N=2$ superconformal coset model based
upon
\ref\KS{Y.~Kazama and H.~Suzuki, \nup{234} (1989) 73.}:
\eqn\hssmodel{{{G_k \times SO_1(dim(G/H))} \over H} \ ,}
where $G/H$ is a hermitian symmetric space.  It was
shown in \NemWar\ that if one topologically twists \ref\twist{E. Witten,
{\it
Commun. Math. Phys.}  {\bf 117} (1988) 353; T. Eguchi, S.-K. Yang, {\it Mod.
Phys. Lett.}  {\bf A4} (1990) 1653.} this model and makes a particular
perturbation (one that also corresponds to a quantum integrable model)
then the result is
the topological $G_k/G_k$ theory.  One now makes the simple observation that
a
$G_k/G_k$ model can be thought of as a $G_k \times G_0/G_k$ model, and one
then
recalls that the lattice construction of $G_k \times G_\ell/G_{k + \ell}$
models is well understood.  The method employed in \MNW\ was thus to follow
through the usual formulation of exactly solvable lattice models as if one
were
going to make a trivial, topological lattice model for $G_k \times G_0/G_k$,
but then make a small modification corresponding to {\it not} imposing the
topological twist and {\it not} enforcing the topological physical state
projection.  This process has only been carried out explicitly for models
with the level, $k$, equal to one, and for $G = SU(N)$.
We will therefore restrict our attention to
such models, but as in the work of \MNW, we do not anticipate any
difficulties in extending our results to other groups or to higher levels.
For this reason, we will, throughout this section, cast our results in as
general a manner as possible, but it should be remembered that all
the details have only been checked completely for $G_k = SU_1(N)$.   We
begin by reviewing the results of \MNW, first in the vertex languange and
then
in the IRF, or height, formulation.  Ultimately we will only use the height
formulation, but we include the vertex description since this was essential
to
the deduction of the proper Boltzmann weights in \MNW.

\nref\PA{V.~Pasquier, \nup{295} (1988) 491. }
\nref\PS{V.~Pasquier and  H.~Saleur, \nup{330} (1990) 523.}
\nref\SaZu{ H.~Saleur and J.-B. Zuber,
``Integrable Lattice Models and Quantum Groups'', in the proceedings
of the 1990 Trieste Spring School on String Theory and Quantum Gravity.}

In the vertex formulation of the models $G_1 \times G_\ell/G_{\ell + 1}$,
one
starts by assigning vectors in the fundamental representation $V$ of $G$
to edges of the lattice (see, for example, \refs{\PA {--} \SaZu}).
The transfer matrix (with free boundary conditions)
is then assembled from products of the $\check R$ matrix of $G$ on $V
\otimes
V$.  This transfer matrix commutes with the action of the quantum group
$U_q(G)$, and so one can quantum group truncate the models to the type II
representations of $U_q(G)$ \refs{\PA {--} \SaZu}.  This can be accomplished
by using the Markov trace in defining the partition function
and correlation functions.  For simply laced groups of rank $r$, the
untruncated theory has a continuum limit corresponding to
$r$ free bosons that have a charge at infinity and that are
compactified on a scaled weight lattice of $G$.  The scale
factor is determined by the $q$-parameter of the quantum group.  After
truncation the model is precisely the desired coset model.

The foregoing is simply the lattice version of the bosonic Coulomb gas
description of these conformal coset models \PS.
It should also be noted that the Hamiltonian of the
system has a boundary term that is precisely the analogue of the charge at
infinity in the continuum Coulomb gas description.  To get the $N=2$
supersymmetric models \hssmodel\
one first twists the $\check R$ matrix so as to modify
this boundary charge to be the one appropriate to the subgroup $H$ of $G$.
This means that the tranfer matrix now only commutes with $U_q(H)$, and so
one can only truncate with respect to $U_q(H)$.  If one performs this
partial
truncation, and sets the level, $\ell$,
equal to zero, then the result is precisely the
model whose continuum limit is $G_1 \times SO(dim(G/H))/H$.  The simplest
way
of understanding this result is to first observe that $H$ must have the same
rank as $G$, and has the form $H= H^\prime \times U(1)$, where $H^\prime$
is semi-simple.  The result of the quantum group truncation with respect
to $U_q(H)$ is simply the model
$H_1 \times H_p/H_{p+1}$.  The level $p$ is equal to  $ \ell + g - h$, where
$g$ is the dual Coxeter number of $G$, and $h$ is the dual Coxeter number of
the appropriate factor of $H$.  (The dual Coxeter number of a $U(1)$ factor
is
defined to be zero.)  Because the states of the model are obtained from
vectors
on the weight lattice  of $G$, this implies that there are very specific
restrictions of, and correlation between, the $H$-weight
vectors in the factors of $H$.  Thus, in the continuum limit, one will
indeed
obtain the $H_1 \times H_p/H_{p+1}$ model, but with a special choice of
modular
invariant.  For $\ell =0$ this special modular invariant is precisely the
one
induced by the conformal embedding $H_{g-h} \hookrightarrow SO(dim(G/H))$.
Similarly, since $H$ and $G$ are simply laced, $H_1$ can be replaced by
$G_1$
and thus the $H_1 \times H_{g-h}/H_{g-h+1}$ model is precisely the $N=2$
superconformal coset model \hssmodel.

\epsfxsize = 3.0in
\vbox{\vskip .3in\hbox{\centerline{\epsffile{latdiag.eps}}}
\vskip .2in
{\leftskip .5in \rightskip .5in \noindent \ninerm \baselineskip=10pt
Figure 1. A section of the lattice upon which the
model is defined.  The bold zig-zag is the initial time slice, and the
arrow indicate orientations of edges.
\smallskip}} \bigskip \bigskip

In the IRF, or height, formulation the foregoing yields a rather simple
prescription.  One first constructs the untruncated models based upon $G$.
The edges of the lattice are oriented, or ordered, along
``space-like'' slices (see Fig. 1).  Heights are assigned to lattice sites
with
the rule that successive heights must differ by some weight, $e_j$, in the
fundamental representation, $V$, of $G$.  For $G = SU(N)$ it is convenient
to introduce an orthonormal basis, $f_j$, $j=1,\dots,N$, in $\IR^N$ and
then the vectors $e_j$ can be written $e_k = f_k -{1\over N}( f_1 + \dots +
f_N)$.  The lattice heights therefore have the form
\eqn\heights{\Lambda ~=~  \Lambda_0 + \sum_{j = 1}^r ~ n_j e_j \ ,}
where $n_j \in \ZZ$ and $\Lambda_0$ is an as yet arbitrary vector.
This ``initial vector,'' $\Lambda_0$,
will play a major role in the forthcoming discussion.

\epsfxsize = 3.0in
\vbox{\vskip .3in \hbox{\centerline{\epsffile{irfdiag.eps}}}
{\vskip .2in \leftskip .5in \rightskip .5in \noindent \ninerm
\baselineskip=10pt Figure 2.
In the IRF formulation the heights, $\Lambda$, are
associated to vertices as shown. \smallskip}}
\bigskip

Consider the
evolution for a typical plaquette as shown in Fig. 2.  Let
$w(\Lambda, \Lambda + e_i, \Lambda + e_j, \Lambda + e_k | u)$
denote the corresponding Boltzmann weight, where $u$ is the spectral
parameter.   For the present, let square brackets around a
variable, $[\nu]$, denote $[\nu] = sin(\pi \gamma \nu)$.  The parameter
$\gamma$ is related to the quantum group parameter by $q = e^{i \pi
\gamma}$.
The critical, and off-critical, Boltzmann weights for models
based on the fundamental representation of any group, $G$, can be found in a
number of references (for example, see
\ref\kyotoa{M.~Jimbo, T.~Miwa and M.~Okado \cmp{116} (1988) 507.}).
For $G = SU(N)$ the non-vanishing \BW are as follows:
\eqn\BWSU{\eqalign{
w(\Lambda, \Lambda + e_i, \Lambda + 2 e_i; & \Lambda + e_i | u) ~=~
{[u+1] \over  [1]} \cr
w(\Lambda, \Lambda + e_i, \Lambda + e_i + e_j; & \Lambda + e_i | u) ~=~
{[(\Lambda + \rho)\cdot(e_i - e_j) ~-~ u] \over
[(\Lambda + \rho)\cdot(e_i - e_j)]} \cr
w(\Lambda, \Lambda + e_i, \Lambda + e_i + e_j; & \Lambda + e_j | u) ~=~ \cr
{[u] \over  [1]} ~ & \left( {[(\Lambda + \rho)\cdot(e_i - e_j) + 1]
[(\Lambda + \rho)\cdot(e_i - e_j) - 1 ] \over
[(\Lambda + \rho)\cdot(e_i - e_j)]^2} \right)^\half \ , }}
where $i \ne j$.
The vector, $\rho$, is the Weyl vector of $G = SU(N)$:
\eqn\weyvec{\rho ~=~ \half \Big( (N-1) e_1 + (N-3) e_2 + \dots \dots
- (N-1) e_N \Big) \ .}
The \BW in \BWSU\ satisfy the Yang-Baxter equations, or, more precisely, the
star-triangle relations (STR) for all values of $\gamma$ and for all
vectors $\Lambda_0$.   The foregoing height model with these \BW is
equivalent
to the untruncated vertex model described earlier.

To obtain the
restricted (RSOS) models corresponding to $G_\ell \times G_1/G_{\ell + 1}$,
one
takes $\Lambda_0 = 0 $, $\gamma = 1/(\ell + g +1)$ and restricts the heights
to
the fundamental affine Weyl chamber of $G_{\ell+1}$.  That is, the heights
must be affine highest weights of $G_{\ell + 1}$.  If one were to make this
choice of $\Lambda_0$ and $\gamma$, but not make this restriction on the
heights, then the \BW would be singular for certain combinations of heights.
It is elementary to verify that the transfer matrix arising from
the \BW in \BWSU\ preserves this restriction.

The corresponding {\it off-critical} \BW are obtained by replacing circular
functions by elliptic functions.  That is, one redefines $[\nu]$ by:
\eqn\ocbws{\eqalign{[\nu] ~\equiv~ &\vartheta_1(\gamma \nu | \tau ) \cr
 ~\equiv~ & 2 p^{1 \over 8} sin(\pi \gamma \nu)~  \prod_{p=1}^\infty
(1 - p^n)(1 - e^{2 \pi i \gamma \nu} p^n) (1-e^{-2 \pi i \gamma \nu} p^n) \
,}}
where $p \equiv e^{2 \pi i \tau}$.  Once again the \BW satisfy the STR for
all
values of $\gamma$, $\tau$ and all vectors $\Lambda_0$.  The RSOS
restrictions
are exactly as described above.

The lattice analogues \MNW\ of the $N=2$ superconformal grassmannian models:
\eqn\grass{{\cal G}_{1,m,n} ~\equiv~ {{SU_1(m+n) \times SO_1(2mn)}
\over {SU_{n+1}(m) \times SU_{m+1}(n) \times U(1) }} \ , }
are constructed by once again taking the heights as in \heights\
(with $N=n+m$).  However one now restricts the heights to be affine
highest weights of both of the
subgroups $SU_{n+1}(m)$  and $SU_{m+1}(n)$ of $SU(m+n)$.  The heights remain
unrestricted in the $U(1)$ direction.  The critical \BW are then as follows:
Consider the evolution shown in Fig. 2.  Let  $\gamma =
{1 \over m+n+1}$, and then\foot{ We have rescaled the \BW of \MNW\ by a
factor of ${1 \over sin(\gamma \pi)}$.}:

\item{(i)} If one has $1 \le i,j \le m$ or $m+1 \le i,j \le m+n$ then
the \BW are exactly those of \BWSU\ with $[\nu] = sin(\pi \gamma \nu)$.

\item{(ii)} If one has $1 \le i,k \le m$ and $ m+1 \le j \le m+n$, or
if one has $1 \le j \le m$ and $ m+1 \le i,k \le m+n$ then one
must have $i=k$ and the Boltzmann weight is $1$.

\item{(iii)}  If one has $1 \le i \le m$ and $ m+1 \le j,k \le m+n$, or
if one has $1 \le j,k \le m$ and $ m+1 \le i \le m+n$ then one
must have $j=k$ and the Boltzmann weight is $sin(\pi u)/sin(\pi \gamma)$.

\noindent It is easy to see that the \BW preserve the partial restrictions
on
the heights.  Such a model should probably be called a partially restricted
solid-on-solid (PRSOS) model.   It is worth noting that the \BW are still
periodic functions of the heights, and so even though the $U(1)$ is
unrestricted, it is compactified at a very particular radius (indeed, at the
``supersymmetric radius'') in the continuum limit.

The problem now is to find elliptic \BW that satisfy the STR, and in the
critical limit, $\tau \to i \infty$, reduce to the foregoing critical
weights.
The solution is almost outrageously simple, and is
in a sense even simpler than the foregoing description of the critical
weights.
One starts by making the partial restriction of the heights as described
above, and one simply uses the \BW given in \BWSU\ with $\gamma = {1 \over
g+1}$ $= {1 \over m+n+1}$.  One cannot take $\Lambda_0 = 0$ since, without
complete restriction in all directions, the evolution would be singular.
However, one simply takes
\eqn\lmbddefn{\Lambda_0 ~=~ {1 \over \gamma g} (\rho_G - \rho_H)~\tau \ ,}
where $\rho_G$ and $\rho_H$ are the Weyl vectors of $G$ and $H$.  For the
grassmannian model \grass, this is simply
\eqn\lmbdgrass{\Lambda_0 ~=~ {m+n+1 \over 2(m+n)}[n(e_1 + \dots + e_m)
- m(e_{m+1} + \dots e_{m+n})] \tau \ .}
Observe that for the physical model, the vector \lmbddefn\
is purely imaginary.  Using this
value of $\Lambda_0$ is the lattice version of ``untwisting'' the
topological energy momentum tensor \foot{Indeed, if the continuum theory
has an extended chiral algebra, then this is the lattice
version of untwisting the extended topological chiral algebra.}.

One can also think of the foregoing models from slightly different
perspective: the heights can be taken on the partially
restricted weight  lattice of $G = SU(m+n)$ with  $\Lambda_0 =0$, but the
\BW must now  incorporate the shift \lmbddefn.  We will henceforth
adopt this perspective. The shift by $\Lambda_0$ does not modify the \BW in
the $H$ direction,
{\it i.e.} there is no change in case (i) above.  The other \BW are,
however, significantly modified.  The effect of this shift is to replace
certain strategic $\vartheta_1$'s by $\vartheta_4$'s,
and this neatly removes all
the potential singularities in the \BW owing to the unrestricted $U(1)$.
In order to describe the results explicitly let $[\nu]$ be given by \ocbws\
and let $\{\nu\}$ is defined by:
\eqn\curly{\eqalign{\{\nu\} ~\equiv~ &\vartheta_4 (\gamma \nu | \tau ) \cr
 ~\equiv~ & \prod_{p=1}^\infty (1 - p^n)(1 - e^{2 \pi i \gamma \nu}
p^{n-\half} )  ~(1-e^{-2 \pi i \gamma \nu} p^{n - \half}) \ .}}
Set the parameter $\gamma$ to ${1 \over m+n+1}$, and take the vector
$\Lambda_0$ to be zero.  The non-vanishing \BW are then as follows:

\noindent If $1 \le i,j \le m$ {\it or} $m+1 \le i,j \le m+n$ then, for
$i \ne j$, one has:
\eqn\ellbwsa{\eqalign{
w(\Lambda, \Lambda + e_i, \Lambda + 2 & e_i; \Lambda + e_i | u) ~=~
{[u+1] \over  [1]} \cr
w(\Lambda, \Lambda + e_i, \Lambda + e_i + & e_j; \Lambda + e_i | u) ~=~
{[(\Lambda + \rho)\cdot(e_i - e_j) ~-~ u] \over
[(\Lambda + \rho)\cdot(e_i - e_j)]} \cr
w(\Lambda, \Lambda + e_i, \Lambda + e_i + & e_j; \Lambda + e_j | u) ~=~ \cr
{[u] \over  [1]} ~ & \left( {[(\Lambda + \rho)\cdot(e_i - e_j) + 1]
[(\Lambda + \rho)\cdot(e_i - e_j) - 1 ] \over
[(\Lambda + \rho)\cdot(e_i - e_j)]^2} \right)^\half \ . }}

\noindent If $1 \le i \le m$ or $m+1 \le j \le m+n$, or vice-versa, then
one has
\eqn\ellbwsb{\eqalign{
w(\Lambda, \Lambda + e_i, \Lambda + & e_i + e_j; \Lambda + e_i | u) ~=~
{\{(\Lambda + \rho)\cdot(e_i - e_j) ~-~ u\} \over
\{(\Lambda + \rho)\cdot(e_i - e_j)\}} \cr
w(\Lambda, \Lambda + e_i, \Lambda + & e_i + e_j;  \Lambda + e_j | u) ~=~ \cr
{[u] \over  [1]} ~ & \left( {\{(\Lambda + \rho)\cdot(e_i - e_j) + 1\}
\{(\Lambda + \rho)\cdot(e_i - e_j) - 1 \} \over
\{(\Lambda + \rho)\cdot(e_i - e_j)\}^2} \right)^\half \ . }}
\nref\kyotob{T.~Miwa, M.~ Jimbo and M.~Okado, ``Symmetric tensors of the
$A^{(1)}_{n-1}$ family,'' Kyoto preprint (1987), in {\it Algebraic
Analysis,}
Festschrift for M.~Sato's 60th birthday, Academic Press (1988). }

It is once again obvious that these \BW preserve the
partial restriction of the heights, and one can easily verify that these
elliptic \BW yield the appropriate critical \BW in the limit $\tau \to i
\infty$.  The fact that they satisfy the STR follows from the original proof
in
\refs{\kyotoa,\kyotob}  where it was shown that the \BW given by \BWSU\ and
\ocbws\ satisfy the STR for all choices of $\Lambda_0$, including complex
values.

Thus, in the height language, off criticality, the lattice analogues of the
$N=2$ supersymmetric quantum integrable models are obtained by making a
model
whose complete restriction would be rigid and related to $G_k \times
G_0/G_{k+0}$, but instead one only makes a partial restriction of the
heights
and then ``topologically untwists'' the model using a uniform shift of the
heights by the vector \lmbddefn.

{}From the Coulomb gas analysis of \MNW\ and as confirmed by the Bethe
Ansatz
computations in \ZMa, the foregoing height model
at criticality yields the Neveu-Schwarz
sector of the $N=2$ superconformal model.  The Ramond sector can be obtained
by
a uniform spectral flow in the $U(1)$ direction.  This is easily
implemented:
one shifts all the lattice height appropriately, or equivalently one takes
$\tau \to \tau + 1$ in \lmbddefn.

\newsec{A simple example}

The simplest example is to take $G = SU(2)$, for which our lattice model
construction collapses to what is basically the IRF version of the eight
vertex model \refs{\Bax {--} \ABF}.   The weight vectors
$e_1$ and $e_2$ of the foregoing section satisfy $e_1 = -e_2$, and the Weyl
vector is given by $\rho = e_1$.  The unrestricted heights lie on
the weight lattice of $SU(2)$, which we will parametrize by an integer,
$a$, where $\Lambda = \Lambda_0 + (a-1) e_1$.  The integer, $(a-1)$,
is equal to twice the spin of the corresponding $SU(2)$ weight.
The  lattice analogues of the
ordinary Virasoro minimal series can be obtained by making the appropriate
quantum group truncation of this $SU(2)$ model \PS.  If one takes
$\Lambda_0 = 0$ and $\gamma = 1/3$ then the quantum group truncated model
is completely rigid with $a$ alternating between $1$ and $2$.
There are thus two states in this model depending upon whether a given site
has height $1$ or $2$.  This is the $SU_1(2)/SU_1(2)$ lattice model.  To get
the lattice analogue of the $N=2$ supersymmetric theory,
one should only quantum group truncate with
respect to the $H$ subgroup, or more precisely, with respect to the
semi-simple part, $H^\prime$, of $H$.  In this instance, this means that
one performs no quantum group truncation at all, and one therefore has a
Gaussian model.  The height, $a$, takes the values $0,1$ or $2$ modulo $3$,
and values of $\gamma$ and $\Lambda_0$ are:
\eqn\params{\gamma ~=~ {1 \over 3} \qquad {\rm and } \qquad
\Lambda_0 ~=~ {3 \over 2}~e_1~\tau \ .}
This value of $\gamma$ means that, in the continuum, this Gaussian model is
compactified on a circle of radius $R = \sqrt{3}$, which is the radius
appropriate the first member of the $N=2$ superconformal minimal series,
with
central charge, $c=1$.  The off-critical $N=2$ supersymmetic,
quantum integrable model corresponds to the most relevant chiral primary
perturbation of the conformal model \FMVW, and may be thought of
as sine-Gordon at the supersymmetric value of the coupling constant.  From
\ellbwsa\ and \ellbwsb\ the elliptic \BW are the following:
\eqn\SGbwsa{\eqalign{
w(a, a \pm 1, a \pm 2; a \pm 1 | u) &~=~ {\vartheta_1({ \pi \over 3}(u+1) |
\tau) \over  \vartheta_1({ \pi \over 3} | \tau)} \cr
 w(a, a \pm 1 , a; a \pm 1 | u) &~=~
{ \vartheta_4({ \pi \over 3}(a  \mp u) | \tau) \over
 \vartheta_4({ \pi \over 3} a | \tau) } \cr
 w(a, a \pm 1 , a; a \mp 1 | u) &~=~ {\vartheta_1({ \pi \over 3} u | \tau)
 \over  \vartheta_1({ \pi \over 3} | \tau) } \left(
 {{ \vartheta_4({ \pi \over 3} (a - 1)
 | \tau)~  \vartheta_4({ \pi \over 3}(a + 1) | \tau) } \over
 (\vartheta_4({ \pi \over 3} a | \tau))^2} \right)^\half \ . }}
\nref\AKTY{A.~Kuniba and T.~Yajima, \jsp{52} (1987) 829.}
\nref\PGins{P.~Ginsparg, \nup{295} (1988) 153.}
\nref\Petal{P.A.~Pierce and K.~Seaton, \annp{193} (1989) 326;
P.A.~Pierce and M.T.~Batchelor, \jsp{60} (1990) 77.}

These \BW are precisely the same as those of the $A_2^{(1)}$ cyclic
solid-on-solid models described in \refs{\Bax,\AKTY {--} \Petal}.  In this
context the labelling of the model by  $A_2^{(1)}$ refers to the fact that,
because of the periodicity of the Boltzmann weights, the unrestricted
$U(1)$ is, in fact, cyclic and so rather than taking the heights to be
in $\ZZ$ one can view them as living on the extended Dynkin diagram of
$A_2$.

\newsec{The Free Energy}

Apart from the fact that it is relatively easy to do, there are several
reasons why it is interesting to compute the free energy per unit volume
for our models. The primary reason is that it provides an indirect
check on the supersymmetry of the model: Because a model with unbroken
supersymmetry has a ground state energy that is equal to zero, one should
expect that, in the supersymmetric regime, the  free energy will be
``essentially zero''.    To be more precise, one should
remember that if all the \BW are multiplied by a uniform, non-vanishing,
analytic function of $u$ and $p$ then they will still satisfy the STR.  As
consequence, the free energy per unit volume for our model can always be
modified by the addition of the logarithm of such an analytic function.
Therefore, the signal of a supersymmetric theory is not that the free
energy vanishes identically, but that it is the logarithm of a non-vanishing
analytic function of $p$ and $u$.  If this is the case then the \BW can then
be modified to make the free energy vanish identically.

This may, at first sight, seem contrary to the fact that these
exactly solvable lattice models undergo second order phase transitions
in the limit as $p \to 0$ since such a transition mandates  non-analytic
behaviour in the thermal parameter as one approaches criticality.
The requisite phase transition will still be present in our models since the
analytic form that we get for the free energy will be retricted to a
particular regime of the model.  The free energy will have other functional
forms in other regimes of the theory.
In the next section we will also show that,
at least in our simple example, the local height probabilities have proper
scaling behaviour as one approaches criticality.

It is certainly not unknown for exactly solvable models to have analytic
free
energies in certain regimes.  It was observed some considerable time ago
\refs{\Baxter,\ABF} that the  eight vertex model has analytic free energy in
regime $I\! I\! I$ when the parameter, $\gamma$ (in our notation), satisfies
$\gamma^{-1} = 2j + 1$, for some positive integer $j$.  The supersymmetric
point is $j=1$.  Analytic free energy therefore does not
imply supersymmetry, however it might imply the existence of
a generalized supersymmetry as in  \ALCV \foot{This suggestion was made
by H.~Saleur.}.  Whether this conjecture
is true or not, it is still an important consistency check to see if the
free
energy per unit volume is analytic for our models.

There is also a practical reason for computing the free energy:
It enables us to relate the elliptic nome, $p$, to the correlation length,
$\xi$, in the scaling region.  Then using hyperscaling arguments and the
corner tranfer matrix we can obtain the scaling properties of various
operators in the theory.

The easiest way to compute the free energy per unit volume, $\scf$,
is to use its analytic and inversion properties (see chapter 13 of
\Baxter).  In this paper we will only consider regime $I\! I\! I$ of the
$SU(N)$ model, {\it i.e.} $ 0 < p \equiv e^{2 \pi i \tau} < 1$,
$-\half N < Re(u) < 0$.  (This regime corresponds to the {\it unitary},
quantum integrable field theory of interest -- other regimes either have
different  conformal limits or correspond to perturbations with imaginary
coupling.)  We will also consider the models defined by \curly\ -- \ellbwsb\
where $\gamma$ is now an arbitrary, positive, real parameter.

To compute $\scf$ it is first convenient to multiply the \BW \ellbwsa\ --
\ellbwsb\ by a factor of
\eqn\invfactor{e^{{i \pi \gamma^2 \over \tau}  (u^2 + {2u \over N})} \ ,}
and perform the modular inversion $\tau \to - 1/\tau$.  After making a gauge
transformation one then finds that the \BW are manifestly periodic under:
\eqn\periodicity{ u ~\to~ u ~+~ {2 \tau \over \gamma}  \ .}
One then writes
$$
\scf ~=~ - \ log(\kappa(u)) \ ,
$$
and requires that $log(\kappa) $ be analytic in a region containing regime
$I\! I\! I$ and also be periodic under \periodicity.  Note that one does
not impose the other periodicity of the theta functions ($ u \to u +
{2 \over \gamma}$).  The periodicity that one imposes is the one that
preserves the regime of interest and that is also manifest in the low
temperature ($\tau \to i0$) limit.

Using the properties of the corner transfer matrix one can show that
$\kappa(u)$ must also satisfy:
\eqn\quadone{ \kappa(u) ~ \kappa(-u) ~=~ h(1-u) ~ h(1+u)  \ ,}
\eqn\quadtwo{ \kappa(\lambda + u) ~ \kappa(\lambda - u) ~=~ h(\lambda - u)
{}~ h(\lambda + u) \ ,}
where $\lambda = -N/2$ for $SU(N)$, and
\eqn\hdefn{h(u) ~\equiv~ {\vartheta_1 ({u \gamma \over \tau} |
{-{1 \over \tau}}) \over \vartheta_1 ({ \gamma \over \tau} |
{-{1 \over \tau}}) } \ .}
The function $h(u)$ that appears on the right hand side of these equations
depends only upon the inversion relations of the elliptic Boltzmann weights.
These equations, along with analyticity and periodicity, determine
$log(\kappa(u))$ completely.  One simply writes $log(\kappa(u))$ as a
general
Fourier series in $e^{i \pi \gamma u/\tau}$ and uses \quadone\ and
\quadtwo\ to determine the coefficients.  This is a little tedious but it
is straightforward.  Alternatively one can also observe that the foregoing
equations and constraints {\it do not depend on the choice of}  $\Lambda_0$.
This means that the free energy for all of of the lattice analogues of the
grassmannian models \grass\ only depends upon $N = m+n$, and this free
energy is exactly that of the topological $SU_1(N) \times SU_0(N)/SU_1(N)$
model.  The fact that the free energy is independent of $\Lambda_0$ also
means that we can use the known results for $\Lambda_0 =0$
for the models based on $SU(N)$ \ref\MPRCAT{M.P.~Richey and C.A.~Tracy,
\jsp{42} (1986) 311.}.  One therefore has
\eqn\freeen{ log(\kappa(u)) ~=~ log \left( {\vartheta_1 (\gamma (u+1) |
\tau) \over \vartheta_1 (\gamma  | \tau) } \right) ~-~
\sum_{k = -\infty}^{k = \infty} ~ f(k; u, \gamma, \tau) \ ,}
where
\eqn\fdefn{ f(x; u, \gamma, \tau) ~\equiv~ {{sinh\big({\pi i \over \tau}
(1 - \gamma) x \big) ~ sinh\big({2 \pi i \over \tau}  \gamma u x \big)~
sinh\big({\pi i (N-1) \over \tau}  \gamma  x \big)} \over
{x ~ sinh\big({\pi i \over \tau} x \big)~ sinh\big({\pi i N \over \tau}
\gamma x \big)}} \ .}
This expression differs slightly from that of \MPRCAT\ in that we have
subtracted the logarithm of the phase \invfactor\ so that the result is
no longer exactly periodic under \periodicity, but so that it does give
the free energy for the model defined by the \BW in \ellbwsa\ and \ellbwsb.

To obtain the result as a function of $p = e^{2 \pi i \tau}$ one needs to
perform the modular inversion of the second term in \freeen.  This can
be done by Poisson resummation (see, for example, chapter 10 of
\Baxter, or appendix D of \ref\kyotoc{E.~Date, M.~Jimbo, A.~Kuniba, T.~Miwa
and M.~Okado, {\it Advanced Studies in Pure Mathematics}
{\bf 16} (1988) 17.}).  That is, one defines
\eqn\fourtrf{ \hat f (\zeta) ~\equiv ~ \int_{-\infty}^{+\infty}~
e^{2 \pi i \zeta x} ~ f(x) ~ dx }
and uses the equality:
\eqn\poiss{ \sum_{k = -\infty}^{k = \infty} ~ f(k) ~=~
\sum_{k = -\infty}^{k = \infty} ~ \hat f(k) \ . }
The fourier transform of $\fdefn$ can be performed by closing the contour
above or below the real axis, depending upon the sign of $\zeta$, and then
summing the residues.  This sum over residues and the sum in \freeen\
generates an expansion in powers of $p$.  The vanishing of
$sinh({\pi i N \over \tau} \gamma x )$ in denominator of \freeen\
gives rise to residues that are proportional to $p^{1 \over N \gamma}$.
This is the source of the non-analytic behaviour of the free
energy as a function of $p$.  The residues coming from the vanishing of the
other $sinh$ function in the denominator of \freeen\ are all proportional
to integral powers of $p$.  It is easy to extract the leading behaviour as
$\tau \to i \infty$ from this sum over residues.  We find
\eqn\freescale{\scf ~\sim~ {{ 4 ~sin \big({1 \over N} \big ({1 \over \gamma}
 -1 \big) \pi \big)  ~ sin \big({2 \pi u \over N } \big)~
sin \big({\pi \over N } \big)} \over
{ sin \big({\pi  \over N \gamma } )} } ~~p^{1 \over N \gamma}  \ .}
For generic values of $\gamma$ this means that $\scf \sim
p^{1 \over N \gamma}$, while from hyperscaling one has $\scf \sim
{1 \over \xi^2}$, where $\xi$ is the correlation length.  Therefore, we have
\eqn\tauxi{ \tau ~\sim~ { i N \gamma \over  \pi} ~ log (\xi) \ ,}
as $\tau \to i \infty$.

One should note that for
\eqn\Fvancond{\gamma^{-1} ~=~ N j ~+~ 1,  \qquad j =1,2,3, \dots}
the numerator of \fdefn\ vanishes whenever $sinh({\pi i N \over \tau}
\gamma x )$ vanishes.  As a result, the Poisson resummation of \freeen\
gives rise to an expansion in integral powers of $p$.  That is, if
$\gamma$ satisfies \Fvancond\ then the free energy is an analytic function
of $p$.  The choice $\gamma^{-1} = N+1 $ corresponds to our supersymmetric
models.

For these special values of $\gamma$ it is, of course, no longer true
that $\scf \sim  p^{1 \over N \gamma}$ as $\tau \to i \infty$.  However,
continuity in $\gamma$ means that the relation \tauxi\ is true even at
these special values.

It turns out that for $\gamma$ given by \Fvancond\ one can easily
express the free energy in terms of theta functions.  Here we will give the
result for the supersymmetric model: $\gamma^{-1} = N+1 $.
The general case is similar. For $\gamma = 1/(N+1)$, the second term in
\freeen\ can be written:
\eqn\stdseries{ {2 \pi i \over \tau}~ u \gamma (1 - 2 \gamma) ~-~
\sum_{k = 1}^{k = \infty} ~ {1 \over k(1 - \tilde p ^k)} ~
(z^k - z^{-k}) ~ (w^{-k} - w^{-kN}) \ ,}
where $\tilde p = e^{-2 \pi i /\tau}$, $z = e^{2 \pi i \gamma u / \tau}$ and
$w = e^{2 \pi i \gamma  / \tau}$.  This is a standard expansion of the
logarithm of the ratio of two theta functions\foot{One simply takes the
logarithm of the product formula for the theta functions, expands all of
the $log(1 -q^n x^{\pm 1})$ terms into power series in $q^n x^{\pm 1}$, and
then  one can perform the sum over $n$ to obtain \stdseries.}.  Finally one
can perform modular inversions on the theta functions and the final result
is
\eqn\finalfree{log(\kappa(u)) ~=~ log \left( {\vartheta_1 (\gamma (u-1) |
\tau ) \over \vartheta_1 (\gamma |  \tau )} \right) ~=~ log \left( {
[u - 1] \over [1] } \right) \ .}

Observe that, as promised, this free energy is analytic in the whole of
regime
$I\! I \! I$.  Indeed, if one multiplies all of the \BW in \ellbwsa\ and
\ellbwsb\ by ${[1] \over [u - 1]}$ then the \BW are still analytic in regime
$I\! I \! I$, and the free energy, $\scf$, vanishes identically.

{}From the foregoing analysis we can also obtain further confirmation of the
fact that the perturbing operator for the continuum quantum integrable
theory
is indeed the supersymmetry preserving perturbation.  We first observe that
the \BW of our models involve $\vartheta_4$, and not just $\vartheta_1$, and
are therefore really only analytic functions of $p^\half$ \foot{Indeed, in
all
the foregoing analysis of $\scf$ we should really have thought in terms of
analyticity in $p^\half$.}.  It is here that we are finally seeing
some, albeit very mild, effect of our choice of $\Lambda_0$.  A different
choice of $\Lambda_0$ would result in different powers of $p$ in the
Boltzmann weights.
It follows that the coupling constant of the perturbing operator, at
least in the scaling region, must be proportional to $p^\half$.  However,
for $\gamma = 1/(N+1)$, \tauxi\ implies that $p^\half \sim
\xi^{-{N \over N+1}}$.  Consequently, the perturbing operator must have
dimension $2 - {N \over N+1}$, or conformal weights:
\eqn\pertwts{h ~=~ \bar h ~=~ \half ~+~ {N \over 2(N+1)} \ .}
These are precisely the weights of the perturbing operators that lead
to the natural quantum integrable, supersymmetric continuum field theories
\refs{\FMVW, \FLMW,\NemWar}.
That is, these conformal weights are those of the operators
\eqn\superperts{\Gminus \widetilde \Gminus \phi \qquad {\rm and} \qquad
\Gplus  \widetilde \Gplus \tilde \phi \ ,}
where $\phi$ and $\tilde \phi$ are the most relevant, chiral primary fields.

More generally, suppose that we take the initial height vector to be
\eqn\newldefn{\Lambda_0 ~=~ {2 \mu \over \gamma g} (\rho_G - \rho_H) ~
\tau \ ,}
where $\mu$ is some parameter.  Then for $0 \le \mu < 1$ the lowest
power of $p$ in the \BW is $p^\mu$.  By the same arguments as above,
this implies that the conformal weights of the perturbing operator are
given by:
\eqn\dimpert{h ~=~ \bar h ~=~ 1 ~-~ {N \mu \over (N+1)} \ .}
In particular, we see that these weights are linear functions of $\mu$,
and that the corresponding operators become marginal in the topological
model.  If we assign an $U(1)$ charge of $-{N \over N+1}$ to the perturbing
operators, then \dimpert\  is precisely consistent with the view that
varying $\mu$ in the initial vector \newldefn\ corresponds to smoothly
untwisting the topological energy momentum tensor according to:
\eqn\untwist{T (z) ~=~ T^{\rm top} (z) ~-~ \mu ~\partial J(z) \ .}
This assignment of charge is also the correct one for the operators in
\superperts.

\newsec{Local Height Probabilities}

Aside from their own intrinsic interest, the computation of local height
probabilities provides further confirmation of our identification of
the continuum limits of these models.  We have, as yet, only performed the
entire computation for the simple model described in section 3.
Indeed the basic computation was already done in \refs{\AKTY,\Petal}.
Before giving a brief discussion of this, and a further refinement of it,
we first describe our expectations based upon preliminary computations
for the general grassmannian models \grass.  Once again we will restrict
our attention to regime  $I\! I \! I$.

\nref\kyotod{E.~Date, M.~Jimbo, A.~Kuniba, T.~Miwa and M.~Okado,
\nup{290} (1987) 231;  M.~Jimbo,  T.~Miwa and M.~Okado,
\nup{300} (1988) 74.}

For the lattice analogues of $SU_1(N) \times SU_\ell(N)/SU_{\ell+1}$ the
local height probabilities were computed in \refs{\kyotoc, \kyotod}.
The probability, $\scp(\Lambda | \nu)$, of having a height $\Lambda$ at the
origin, given that there is some ground state,  $\nu$, at infinity has the
generic form
\eqn\lhp{\scp (\Lambda | \nu) ~=~ b_{\Lambda, \nu} (x^N) ~ u_\Lambda(x) ~
V_\nu(x) \ ,}
where $x = e^{-{{2 \pi i} \over {(\ell + N+1) \tau}}}$.  The function
$b_{\Lambda, \nu}$ is a particular branching function of the original
coset model and the function $u_\Lambda(x)$ is an easily computable
combination\foot{ The precise form is determined by the inversion relation
and the gauge transformations used in the simplification of the zero
temperature limit  of the corner transfer matrix.}
of theta functions and an overall power of $x$.
The function $V_\nu(x)$ is determined solely by the
requirement that the local height probabilities are properly normalized,
{\it i.e.} $\sum_\Lambda \scp (\Lambda | \nu) ~=~ 1$.  The heart of the
matter
is thus the branching function $b_{\Lambda, \nu}$.

One establishes the foregoing result by showing
that the corner transfer matrix
essentially has integer spectrum and then computes this spectrum in the
zero temperature limit \refs{\Baxter, \ABF}.  The resulting expressions
for the (non-normalized) local height probabilities are then
obtained in terms of one-dimensional configuration sums.
These configuration sums are taken over walks on the set of allowed
heights for the RSOS model.  If, however, one does not restrict the heights
({\it i.e.} one uses the unrestricted heights of the corresponding SOS
model) one can show that the resulting configuration sum yields a gaussian
partition function.    In a manner rather reminiscent of the computation
of the branching functions of the coset, one can implement the height
restriction by making an additional alternating sum over the Weyl group
of $SU(N)$, and the result is indeed one of the requisite branching
functions.

Based on the details of the computation of the local height probabilities
in the  $SU_1(N) \times SU_\ell(N)/SU_{\ell+1}$ model, we think it extremely
plausible that a partial restriction of the heights will result in the
corresponding partial sum ({\it i.e.} over the Weyl group of $H$ in
\hssmodel) of gaussian partition functions.  It was shown in \MNW\ that
if one takes $\ell = 0$ then such partial sums do indeed yield the
characters of the underlying $N=2$ superconformal coset models.  Clearly
there is much to be verified here, and there are one or two minor subtleties
that we have already encountered.  The complete computation for the general
grassmannian models is in progress and will be published elsewhere
\ref\FutNW{D.~Nemeschansky and N.P.~Warner, to appear.}.  Here we summarize
the local height probabilities for $G=SU(2)$, the details may be found in
\refs{\AKTY,\FutNW}.

We will work entirely in the Neveu-Schwarz sector of the theory.  Because of
the cyclicity of the model, there are only three independent heights, which
we will label by $0,1$ or $2$ taken modulo $3$.   The ground states have an
atypical structure in that one expects, but does not quite find, that
they consist of heights alternating
between some height $b$ and another height $c=b \pm 1$ on neighbouring
lattice
sites.   One can see this by taking the zero temperature limit, and one
finds
that, contrary to the usual experience, the one dimensional configurations
do not exactly diagonalize the hamiltonian of
the corner transfer matrix.  Indeed, this hamiltonian
is not diagonal on configurations involving the height sequence
$(\dots, 0, \pm 1, 0, \dots)$, however, the hamiltonian
does diagonalize on the sums and
differences of such configurations, {\it i.e.} on the combinations:
$(\dots, 0, +1, 0, \dots ) \pm (\dots, 0,  -1, 0, \dots)$.  The symmetric
combination has lower energy than the skew-symmetric combination.   One can
then consistently relabel the symmetric combination by the sequence
$(\dots, 0,  1', 0, \dots)$ and the skew-symmetric combination by
$(\dots, 0, (-1)', 0, \dots)$, and therefore we will henceforth drop the
primes on these height labels.  The effect of this is that for the
ground states the values of $(b,c)$ are $(0,1)$, $(1,0)$, $(1,2)$ or
$(2,1)$,
but {\it neither} $(0,-1)$ {\it nor} $(-1,0)$.

The result of the one dimensional configuration sum with a height $a$ at
the origin and a ground state $(b,c)$ at infinity is:
\eqn\configsum{b_{a,(b,c)} ~=~ {1 \over \eta(q)} ~ \sum_{n = -
\infty}^\infty~
q^{{3 \over 2} \left(n + {a \over 3} - {1 \over 4}(b+c-1) \right)^2 +
\Delta } \ ,}
where
$$
\Delta ~\equiv~ {a^2 \over 12} ~+~ {1 \over 4}(bc - a) -{3 \over 32}
(b+c-1)^2 + {1 \over 24} \ , \qquad
\eta(q) ~\equiv~ q^{1 \over 24} ~ \prod_{n=1}^\infty ( 1 ~-~ q^n) \ ,
$$
and
\eqn\qdefn{ q ~\equiv~ e^{-{4 \pi i \over 3 \tau}} \ .}

The function $b_{a,(b,c)}$ can be thought of as a branching function of
$SU(2) \times SO(2)/U(1)$, and is indeed a character of the $N=2$
superconformal minimal model with $c=1$.
The functions $u_a$ and $V_{(b,c)}$ are given by:
\eqn\uVres{\eqalign{u_a(\tau) &~\equiv~ q^{-{1 \over 12} (a^2 - 3a)}
\vartheta_4(\coeff{a}{3}|\tau) \cr V_{(b,c)}(\tau) &~\equiv~
\sum_{a=0}^2~ u_a (\tau)~ b_{a,(b,c)} \ .}}
Note that the factor of $q^{-{1 \over 12} (a^2 - 3a)}$ in $u_a$ cancels
the $a$-dependence in $\Delta$.
The local height probability is then given by \lhp.

To obtain the critical limit ($\tau \to i \infty$) of \lhp\ one
performs the modular inversion on the part of \lhp\ that is written in
terms of $q$.  One then obtains:
\eqn\Pnice{\scp(a|b,c) ~=~
{{\vartheta_4 (\coeff{a}{3} | \tau) ~ \chi (a| b,c)} \over
{ \eta\big(p^{3 \over 2}\big) \left( \sum_{d=0}^2
\vartheta_4 (\coeff{d}{3} | \tau) ~\chi (d| b,c) \right)}}\ . }
where
\eqn\chidefn{ \chi (a| b,c) ~\equiv~  \sum_{n = - \infty}^\infty~
e^{{i \over 2} n^2 \tau + 2 \pi i n ( {a \over 3} + {b+c-1\over 4})}\ .}
First observe that as $\tau \to i \infty$ one has the expected result that
$\scp(a| b,c) \to {1 \over 3}$, independent of $a$, $b$ and $c$.
There are three natural order parameters defined by:
\eqn\ordpar{\Phi_j ~\equiv~ {1 \over 3} ~ \sum_{a=0}^2 e^{2 \pi i a j / 3}
{}~\scp(a | b,c) \ ,}
for $j =0,1$ and $2$.  As $\tau \to i \infty$ these order parameters scale
as $p^0$, $p^{1 \over 4}$ and $p^{1 \over 4}$ respectively.

To get the physical dimensions of these operators one uses \tauxi\ with
$N=2$ and $\gamma = {1 \over 3}$.  One then obtains:
\eqn\thistauxi{\tau ~\sim~ {2 i \over 3 \pi} ~ log (\xi) \ ,}
which is consistent with the results of \AKTY.  We therefore find that
the scaling dimensions of the $\Phi_j$ are $0$ for
$j=0$ and ${1 \over 3}$ for $j = 1,2$.  This is consistent with the
identification of the continuum, critical limit with the $c=1$, $N=2$
superconformal minimal model and the order parameters are simply the
identity operator and the chiral and anti-chiral primary fields
respectively.

Finally, for the lattice analogues of $N=2$ supersymmetric models, one
can make a further refinement of the corner transfer matrix computation
of local height probabilities, or one-point functions.  Since these models
have an unrestricted $U(1)$ direction, and an associated $U(1)$ charge,
one can keep track of this quantum number in the computation of the
one-point functions, and in particular count the winding number in
the $U(1)$ direction as a particular lattice configuration interpolates
between the height $a$ at the origin and the ground state $(b,c)$ at
infinity.  This means that one computes:
\eqn\onepfn{\scq(a|b,c)(\tau,\nu) ~=~ {{Tr_{(b,c)}(S_a ABCD
e^{2 \pi i \nu J_0})} \over {Tr_{(b,c)}( ABCD )} } \ ,}
where, as usual, $A,B,C$ and $D$ are the corner transfer matrices in
each quadrant,
$S_a$ is the projection onto states with height $a$ at the origin,
$Tr_{(b,c)}$ denotes the trace with boundary conditions $(b,c)$ at infinity,
and the operator $J_0$ counts the $U(1)$ height difference (including
winding
number) between the origin and the ground state at infinity.  The
computation
of this one-point function is a straightforward modification of the usual
one.
The one dimensional configuration sums now get a further weighting by
$e^{2 \pi i \nu}$ to the power of the
winding number, and the net effect is to introduce a further factor of
$e^{2 \pi i (n + {a \over 3} - {1 \over 4}(b+c-1)) \nu}$ into the
summand of \configsum.  This means that the one-point function \onepfn\
is now given by \lhp\ where $b_{a,(b,c)}$ is the complete $N=2$
superconformal character that is further refined by the $U(1)$
quantum numbers.  The additional factor, $u_a$, is the same as it
was in the computation of the local height probability, while the
role of the factor $V_{(b,c)}$ is less clear as this was introduced
to normalize the probability, and this new one-point function does not
appear to have a probabilistic interpretation.

If one sets $\nu = \half$
then this one point function corresponds to the insertion of $(-1)^F$,
where $F$ is the fermion number, along the the positive real axis.  Based
on experience with the continuum, critical limit and the fact that the
punctured plane can be mapped to the cylinder, one might hope that
\onepfn\ would reduce to some form of index.  The refinement of
$b_{a,(b,c)}$
certainly has the requisite properties (for the ground states $(1,2)$ and
$(2,1)$) given that it is already a superconformal character, however
the functions $u_a$ and $V_{(b,c)}$ do not depend upon $\nu$ and are thus
unchanged.  In the continuum, critical limit ($\tau \to i \infty$)
the foregoing one-point function does, of course, reduce to the usual index
computation and the functions $\vartheta_4$ in $u_a$ and
$V_{(b,c)}$ tend to $1$.

We anticipate that the generalization of \onepfn\ to arbitary lattice
analogues
of $N=2$ supersymmetric models will also give rise to configuration
sums that yield the complete $N=2$ superconformal characters, refined by
the $U(1)$ charge. We also expect that in the limit $\tau \to i \infty$,
these one-point functions will yield the corresponding index and even the
generalized index \ref\harv{S.~Cecotti, P.~Fendley, K.~Intriligator and
C.~Vafa, \nup{386} (1992) 405.} at criticality.   The fact that the
configuration sums for these one-point functions lead to the complete $N=2$
superconformal characters appears, at present, to be limited to an amusing
mathematical result.  It would be nice to find a physical interpretation for
the complete character in this context, and not just its $\tau \to i \infty$
limit.

We had originally hoped\foot{These comments are based on a number of
discussions with H.~Saleur.} that \onepfn\ might have given an explicit
formula for the generalized index of the massive $N=2$ supersymmetric
quantum integrable models.  One can differentiate $\scq(a|b,c)(\tau,\nu)$
with respect to $\nu$ and then set $\nu = \half$.  The result corresponds
to a two-point function between a fixed height at the origin and one of the
ground states at infinity, with an insertion of $F (-1)^F$ along the
positive real axis.  This function has a highly non-trivial dependence on
the length scale, $\xi$; and, by conformal invariance, it reduces to the
generalized index in the continuum at criticality.  The hope was that it
would be a solution to some for of topological/anti-topological fusion
equation \ref\SCCV{S.~Cecotti and C.~Vafa, \nup{367} (1991) 359.} in the
continuum limit.  The problem is that the continuum limit requires that one
send $\xi \to \infty$ ({\it i.e.} $\tau \to \infty$), and thus the
function collapses to a number. (This number is, of course, the
generalized index at the conformal point.)  Therefore, this simple
corner transfer matrix computation does not appear to yield the functional
dependence of the generalized index upon the off-critical coupling.  To
obtain this from our \BW one will probably have to do a Bethe Ansatz
computation on a cylinder.

\newsec{Conclusions}

We think that it is now fairly evident that we have elliptic \BW
for models whose continuum limits are a massive, supersymmetry preserving,
quantum integrable perturbations of the $N=2$ superconformal coset models
based on hermitian symmetric spaces.

We also think that there is some
evidence that the the lattice model knows about the supersymmetry.
It was, however,  an adage of
Richard Feynmann that one only advances several arguments as to why a
thing is true when one does not have one {\it good} argument.
The good argument that is lacking here is
the fact that the supersymmetry is not explicitly realized.
One would like to find lattice operators
whose anti-commutator algebra generates the hamiltonian of the transfer
matrix.  While the free energy computation and the links with the
topological $G/G$ structure hints at $N=2$
supersymmetry on the lattice Hilbert space, we do not know of any
operators whose algebra might be interpretted as supersymmetry.
It should also be emphasized that in this paper we have made no direct use
of supersymmetry in performing computations.  On the contrary, the
corner transfer matrix and free energy were derived  using entirely
standard methods, and the results were used as evidence for supersymmetry.
Based upon experience with the \LG approach in the continuum, one would hope
that if one could find  $N=2$ supersymmetry on the lattice, then one would
be able to exploit it directly, and reformulate the lattice theory in
such a manner that the supersymmetry is manifest and certain computations
dramatically simplify.  Thus finding such a lattice supersymmetry is
highly desirable.

There are a number of ways in which one might try to formulate models
that might possibly yield manifest lattice supersymmetry.
For example, one could try to find exactly solvable models with
bosonic and fermionic
variables on the lattice and which explicitly realize the superalegbra on
the
lattice variables.  However, it is not clear
that even this will guarantee global
supercharges with the proper anti-commutators.  Another hope is that
the supersymmetry algebra will appear as a natural extension of the
the underlying quantum group to some larger, affine quantum group.
For our models, the operators that extend $U_q(H)$ to $U_q(\widehat G)$
are not the supersymmetries, but are the relevant perturbations
that lead to a massive quantum integrable model.  This is because the
quantum group that is realized on the lattice is the one associated to the
denominator factor of $H_1 \times H_{g-h}/H_{g-h+1}$.  The supersymmetry
generators are the extension of the quantum group associated with the
numerator factor, $H_{g-h}$ (see \refs{\NemWar,\MNW} for details).  (This
fact
was exploited extensively in \ref\LNW{A.~LeClair, D.~Nemeschansky and
N.P.~Warner \nup{390} (1993) 653.}, where this quantum group plays a
major role in the scattering matrices of solitons of the quantum integrable
model.) Thus, if one is seeking the supersymmetry as part of an affine
quantum
group then our lattice formulation is based on the wrong quantum group.  For
the $N=2$ supersymmetric, minimal model with $c=1$ it is known how to make
a lattice formulation based on the affine $U_q (\widehat{SU(2)})$ where
the quantum group generators appear to be the supersymmetry generators
\ref\Sal{H.~Saleur,  unpublished.}.  However it is not
clear how to generalize this to higher groups, and even in the known
$c=1$ model the putative supersymmetry algebra has rather trivial
anti-commutators, that certainly do not give the spin-chain hamiltonians
\Sal.    Therefore finding manifestly supersymmetric,
exactly solvable lattice models is still very much an open problem.

\nref\PFKI{P.~Fendley and K.~Intriligator, \nup{372} (1992) 533; \nup{380}
(1992) 265.}
There are two  other
implicit features of our lattice models that we wish to bring
out into the open.  First, the fact that the \BW can be normalized so that
the free energy per unit volume is zero means that the corner transfer
matrix does not need to be scaled in order to yield a finite result
in the infinite lattice limit.  This may well afford the interesting
possibility of making physically meaningful comparisons between local
height probabilities with distinct ground states at infinity.

The second feature relates to the fact that most of the  $N=2$
superconformal coset models are known to have two,
or more, $N=2$ supersymmetry preserving perturbations that lead to quantum
integrable models.  In the continuum the easiest way to show this is to use
the fact that
$N=2$ superconformal models have two or more formulations related to
Toda, or para-Toda theories, and then make the affine extensions of these
models \refs{\FLMW,\NemWar}.  Equivalently, there are often two, or more,
coset models that give rise to the same $N=2$ superconformal
model, and for each such coset formulation there is a natural, and distinct
quantum integrable model.  For example, the $N=2$ superconformal minimal
models
with $c = {3k \over k+2}$ can be thought of as either
$SU_k(2) \times$ $SO(2)/U(1)$ or $SU_1(k+1) \times$
$SO(2k)/(SU_2(k) \times U(1))$, however the corresponding quantum integrable
models are very different (see, for example, \refs{\PFKI,\NemWar,\LNW}).
The situation for our lattice models directly parallels this: a given $N=2$
superconformal model has two or more distinct critical lattice
formulations, however, the off-critical versions of these distinct lattice
formulations will lead to the distinctly different quantum integrable
models.

Finally, while we have concentrated on $N=2$ supersymmetric models
with $G= SU_1(N)$, it is fairly clear that our method generalizes.
First, we anticipate little, or no, difficulty in making models based on
other
groups, and generalizing to fused versions of our models.  We expect that
the
prescription will be exactly the same.  There are also obvious
generalizations
in which one once again does not restrict in the $U(1)$ direction, one
chooses
$\Lambda_0$ exactly as we did here, but then does {\it not} set
$\ell = 0$ in $\gamma = 1/(\ell+g+1)$.
The result will be some special modular invariant of the $H \times H/H$
model,
but not at the supersymmetric radius.  (As we have already remarked,
some of these will probably correspond to theories will the generalized
supersymmetry of \ALCV.)   We also expect that one should be able to
get around the restriction to hermitian symmetric spaces.  One might also be
able to construct models in which the heights are unrestricted in more than
one $U(1)$ factor.
One other class of models that will be of interest, but for
which it is less clear how to obtain them, are the lattice analogues
of the $N=1$ and $N=4$ coset models.  One can certainly perform the
appropriate partial restrictions, but one does not have an obvious analogue
of untwisting the \BW using the vector $\Lambda_0$.  At any rate, we wish
to note that we certainly have not constructed the most general PRSOS
models,
and there almost certainly some other interesting ones that we have
not described  here.

\medskip

\noindent
{\bf Acknowledgements}

\smallskip

We would like to thank H.~Saleur for extensive discussion and education on
the
subject of lattice models, and most particularly on the details of
the corner transfer matrix.  We are also grateful for his comments
the manuscript. This work supported in part by funds
provided by the DOE under grant No. DE-FG03-84ER40168.  N.W. is grateful
to the ITP in Santa Barbara for hospitality while this
work neared completion, and
as a result this work was also supported in part by the National Science
Foundation under Grant No. PHY89-04035.

\vfill
\listrefs
\eject
\end